\documentclass[conference]{IEEEtran}

\usepackage{cite,url}
\usepackage{graphicx}
\usepackage{float}
\usepackage[caption=false]{subfig} 
\usepackage{subfig}

\usepackage{amsmath}
\usepackage{amsfonts}
\usepackage{array}

\usepackage{gensymb}

\newcolumntype{M}[1]{>{\centering}m{#1}}
\newcolumntype{N}{@{}m{0pt}@{}}
\begin{document}
\title{How Close Can I Be? - A Comprehensive Analysis of Cellular Interference on ATC Radar}
\author{\IEEEauthorblockN{
 Neelakantan Nurani Krishnan\IEEEauthorrefmark{1},
  Ratnesh Kumbhkar\IEEEauthorrefmark{1},
Narayan B. Mandayam\IEEEauthorrefmark{1}, 
Ivan Seskar\IEEEauthorrefmark{1} and
Sastry Kompella\IEEEauthorrefmark{2}}
\IEEEauthorblockA{\IEEEauthorrefmark{1}WINLAB, Rutgers, The State University of New Jersey, North Brunswick, NJ, USA.\\ Email: \{neel45, ratnesh, narayan, seskar\}@winlab.rutgers.edu}
\IEEEauthorblockA{\IEEEauthorrefmark{2}Information Technology Division, 
Naval Research Laboratory, Washington DC, USA\\
Email: sk@ieee.org}}

\maketitle

\begin{abstract}
Increasing data traffic demands over wireless spectrum have necessitated spectrum sharing and coexistence between heterogeneous systems such as radar and cellular communications systems. 
In this context, we specifically investigate the co-channel coexistence between an air traffic control (ATC) radar and a wide area cellular communication (comms) system. We present a comprehensive characterization and analysis of interference caused by the comms system on the ATC radar with respect to multiple parameters such as radar range, protection radius around the radar, and radar antenna elevation angle. 
The analysis suggests that maintaining a protection radius of $50$ km around the radar will ensure the required INR protection criterion of $-10$ dB at the radar receiver with $\sim 0.9$ probability, even when the radar beam is in the same horizon as the comms BS. Detailed evaluations of the radar target detection performance provide a framework to choose appropriate protection radii around the radar to meet specific performance requirements. 
\end{abstract}


\section{INTRODUCTION} 
\label{sec:intro}
The amount of data traffic carried over wireless spectrum has been increasing over the last decade with the advent of many new applications. It has been estimated that mobile data traffic, which heavily relies on wireless spectrum for connectivity, will observe a sevenfold increase between $2016$ and $2021$ \cite{ciscotraffic}. These traffic requirements have pushed the research community to enable more efficient usage of the already available spectrum and have urged regulatory bodies such as the Federal Communications Commission (FCC) to open newer unexplored bands. 
To implement efficient utilization of available spectrum, many advancements have been made in the fields of cognitive radio systems\cite{mitola99cognet,haykin05cognetbrain}, dynamic spectrum sharing \cite{peha05specshare}, efficient waveform design \cite{schaich2014waveform}, and coexistence between heterogeneous systems \cite{zhang2015lte, khawar2014spectrum}. In an effort to provide more resources, the FCC has not only opened newer bands in the mmWave range for licensed and unlicensed access \cite{niu2015survey} but also some bands for commercial use which are commonly occupied by federal, military or satellite communication systems. Examples of such bands are the TV white space channels between $470$ to $698$~MHz and citizen broadband radio service (CBRS) bands at $3.5$~GHz.  However, the commercial usage of these bands is restricted and governed by policies set by the regulatory bodies due to the presence of incumbent primary users. In order to protect the primary transmissions in these bands from incoming secondary transmissions, the following three approaches are commonly used. The first method is the use of a geolocation database which contains information about the availability of spectrum for secondary usage.  The second approach is  the use of a tiered system where a secondary user detects the presence of a primary transmission using spectrum sensing and utilizes the spectrum only in absence of primary user. However, when the primary system is of critical nature such as a military or an air traffic control (ATC) radar, a third approach is used in which the regulatory body declares a protection region where no secondary transmission is allowed. The choice of this protection region is important and critical for both primary and secondary systems. A small protection region may result in an unacceptable interference to the primary system, while a large protection region may result in an unnecessary loss of transmission opportunity for the secondary system.

In the context of the aforementioned third strategy, this paper investigates the co-channel coexistence of an air traffic control (ATC) radar, operating in the frequency range of $2700$--$2900$~MHz, with a cellular communication (comms) system. The comms base stations are deployed in such a manner that a protection region is maintained around the radar under consideration. A comprehensive analysis of the aggregate interference caused by the comms system on the ATC radar is presented with respect to multiple parameters such as radar target range, protection radius around the radar, and radar antenna elevation angle. This analysis is further utilized to present a framework which can determine a suitable protection radius while meeting any given radar performance requirement.

\subsection{Related Work}
The study of coexistence between radar and other commercial communication systems has been of great interest in the research community and this field has been explored with different variants of radars \cite{khawar2014spectrum, rahman2011feasibility, wang2015atcradar, hessar2016radar}. Authors in \cite{khawar2014spectrum} propose modification in  radar signals in order to avoid interference to a chosen LTE base station. This method requires significant cooperation between these two vastly different systems and may not be a practical solution for tactical systems where information cannot be shared. In \cite{rahman2011feasibility} and \cite{wang2015atcradar}, the authors specifically study the coexistence between ATC radars and LTE, and provide a mutual interference analysis to determine an appropriate protection region around the radar. However, these studies only consider adjacent channel coexistence and  effects of interference on the radar due to a single base station.
In \cite{hessar2016radar}, the authors consider the coexistence between a search radar ($2700$--$2900$~MHz) and WLAN systems, and provide the analysis to find a protection region around radar for single and multiple interferers. However, the authors in this study use only a fixed pathloss model between radar and base stations and do not consider the effects of fading.

This paper differs from existing literature in the following aspects --- (a) a comprehensive co-channel interference characterization and analysis at the radar is presented in the presence of multiple comms base stations, (b) large scale fading coupled with appropriate modeling of pathloss is considered to model the channel between radar and base stations (to account for the stochastic nature of the channel), and (c) accurate radar beam pattern as specified by regulatory bodies is used to characterize the aggregate interference at the radar receiver holistically. 

We note here that a similar aggregate interference analysis can be performed for any type of radar (ATC, naval, military). The choice of ATC radar for the current analysis is primarily due to the fact that the specifications (as mentioned in Table~\ref{specifications}) for an ATC radar are publicly accessible. We also believe that the results from this analysis will serve as a motivation for regulatory bodies to open up bands in the $2700$--$2900$~MHz band for potential unlicensed secondary usage.  

\section{COEXISTENCE SYSTEM MODEL} 
\label{radar-model}
\begin{figure}[t]
\centering 
\includegraphics[width=0.38\textwidth]{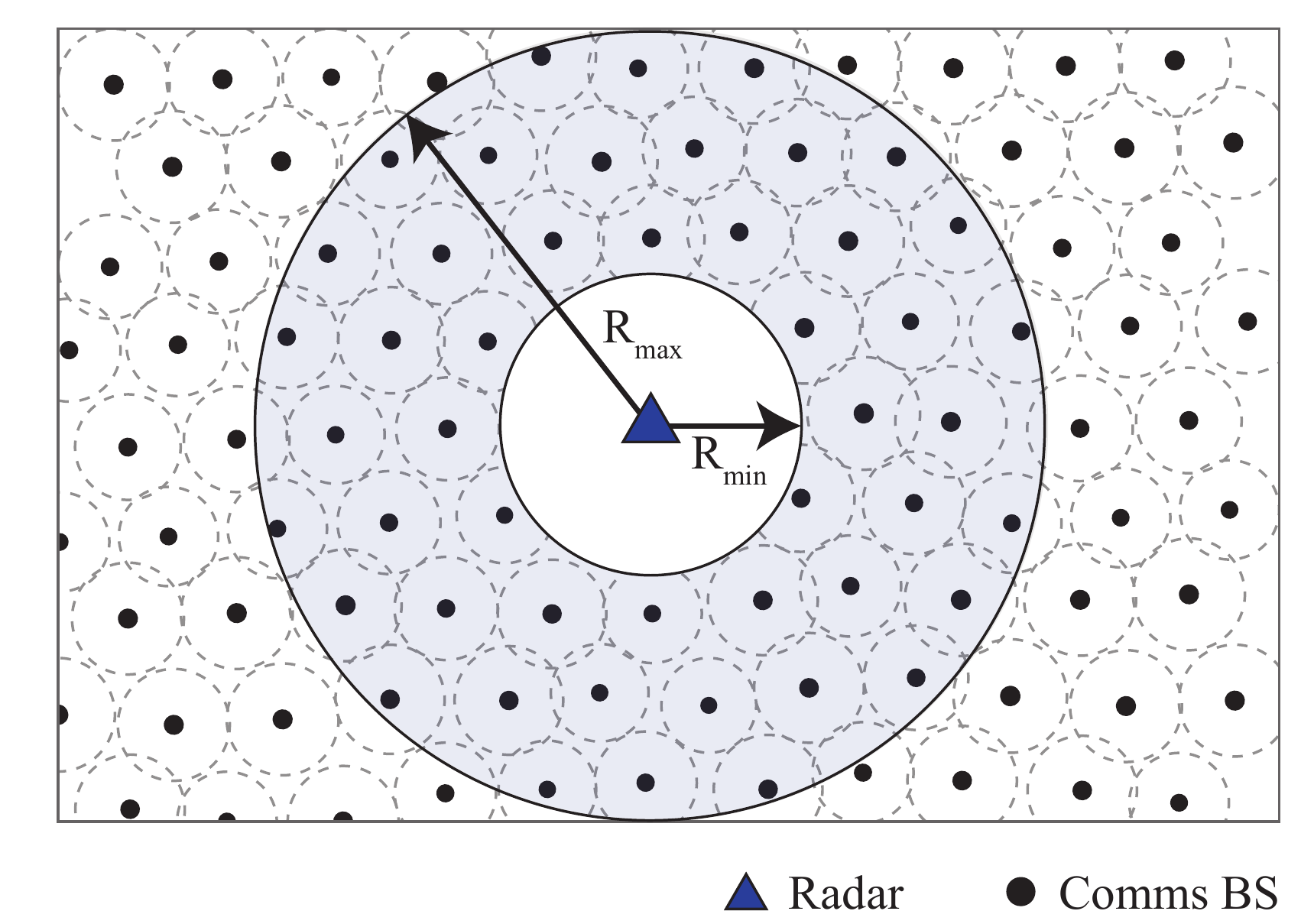}
\caption{Comms BS deployed in an annular region across the radar of interest; note the protection radius around the radar}
\label{scenario}
\end{figure}

This paper explores the feasibility of co-channel coexistence of a wide-area cellular communication systems (comms) with an ATC radar of Type-B \cite{itu2015atcradar}, generally located in airports. A typical coexistence scenario between the two systems is shown in Fig.~\ref{scenario}. Since it is imperative that the performance of the ATC radar is not compromised at any cost, comms base stations (BS) are deployed beyond a certain distance from the radar known as the protection radius ($\text{R}_\text{min}$). Distance beyond which the comms BS have no significant impact on the radar performance is denoted as $\text{R}_\text{max}$.
The analysis in this paper characterizes the impact of protection radius on preserving the radar target detection performance, and provides a framework to answer the question - how close can the comms BS get to the radar before the latter's performance starts deteriorating beyond acceptable limits? 

This work considers a ground-based ATC radar of Type-B with operating parameters as specified in Table~\ref{specifications}. The radar uses a parabolic reflector and has a horizontal scan of $360^{\circ}$ (it does not have a vertical scan). 
The specifications of coexisting comms BS and channel modeling are listed in Table~\ref{specs_LTE}. The following system model and associated analysis assume that the interference from comms BS to radar is primarily from downlink transmissions in the same channel as the radar.
\begin{table}[t]
\centering
\caption{ATC Type-B Radar Specifications}
\begin{tabular}{|M{0.2\textwidth}|M{0.07\textwidth}|c|}
\hline 
\textbf{Operating Parameters} & \textbf{Notation} & \textbf{Value} \\ \hline \hline 
Center Frequency & $f$ & $2900$ MHz \\ 
Wavelength & $\lambda$ & $0.103$ m \\
Transmit Power & $\text{P}_\text{S}$ & $1.32$ MW \\ 
Noise Power & $\text{P}_\text{N}$ & 10$\log$(kTB) + NF dB\\
Boltzmann's constant & k & $1.38 \text{x} 10^{-23}$ J/K \\ 
Temperature & T & $290$ K \\ 
Channel Bandwidth & B & $10$ MHz \\ 
Receiver Noise Figure & NF & $3$ dB \\ 
Radar Cross Section & $\Omega$ & $100 \, \text{m}^2$ \\ 
Interference-to-Noise Ratio & I/N & $-10$ dB \\ \hline
\textbf{Antenna Parameters} & & \\
Radiation Pattern & $\text{g}(\phi)$ & ITU-R M.1851 \\
Pattern Type & - & Cosecant-squared \\
Main Beam Gain & $\text{G}_\text{T}$ & $33.5$ dBi \\
3-dB Beamwidth (Azimuth) & - & $2^\circ$ \\ 
3-dB Beamwidth (Elevation) & - & $4.5^\circ$ \\ 
Side Lobe Level & - & 7.3 dBi \\
Antenna Height & $\text{h}_\text{R}$ & $8$ m \\ \hline 
\end{tabular}
\label{specifications}
\end{table} 

\begin{table}[t]
\centering 
\caption{Comms BS Specifications and Channel Modeling}
\begin{tabular}{|M{0.21\textwidth}|c|c|}
\hline
\textbf{Operating Parameters} & \textbf{Notation} & \textbf{Value} \\ \hline \hline 
Transmit Power & $\text{P}_\text{L}$ & $30$ dBm EIRP \\
Antenna Height & $\text{h}_\text{BS}$ & $30$ m \\
Channel Bandwidth & B & $10$ MHz \\
Pathloss Model & $\rho$ & Extended HATA Model \\
Environment & - & Outdoor Urban \\
Fading & F & Log-normally distributed \\ 
Std deviation of fading & $\sigma$ & $8$ dB \\ \hline 
\end{tabular}
\label{specs_LTE} 
\end{table} 
\section{AGGREGATE INTERFERENCE MODELING}
\label{sec:interf:main}
The modeling of interference from comms BS to radar requires attention due to large distance (typically in orders of tens of kilometers) between the two. 
Owing to this large distance, the propagation environment and received signal strength will be predominantly impacted by pathloss and large-scale fading due to shadowing from obstacles, and not by small-scale fading. The following subsections describe the pathloss and fading models used in the current analysis followed by an analytical characterization of the aggregate comms interference at the radar receiver. 

\subsection{Pathloss and fading Model} 
\label{sec:interf:pathloss}
\begin{table}[t]
\centering 
\caption{Range of Valid Parameters for eHATA Model}
\begin{tabular}{|c|c|} 
\hline
\textbf{Quantity} & \textbf{Valid Range} \\ \hline \hline 
Center Frequency & $1500\, \text{MHz}-3000\, \text{MHz}$  \\
Link Distance & $1\, \text{km}-100\, \text{km}$ \\
Transmitter Height & $30 \, \text{m}-200 \, \text{m}$\\
Receiver Height & $1 \, \text{m}-10 \, \text{m}$  \\ \hline 
\end{tabular}
\label{eHATA} 
\end{table} 
This work uses the extended HATA (eHATA) model in point-to-point mode to compute the median transmission loss between a comms BS and the intended radar receiver. The eHATA model is valid for the set of parameters mentioned in Table~\ref{eHATA}. The comms BS are assumed to be deployed in an outdoor urban region. The median pathloss for a link of distance $r$ km is determined by the following set of equations \cite{eHATA} --- 
\begin{align*} 
\rho\text{(dB)} =& 30.52 + 16.81\log(f) + 4.45\log(f)^2 + \\ &(24.9 - 6.55\log(\text{h}_\text{BS}))\log(\text{R}_\text{bp}) + 10n\log\left(\frac{r}{\text{R}_\text{bp}}\right) \\ &+ 13.82\log\left(\frac{200}{\text{h}_\text{BS}}\right) + a(3) - a(\text{h}_\text{R}) + \text{FSL}(f,\text{R})
\end{align*}
where 
\begin{itemize} 
\item $\text{R}_\text{bp}$ is the break point distance 
\begin{equation*} 
\text{R}_\text{bp} = \left(10^{2n_h}\frac{a_\text{bm}(f,1)}{a_\text{bm}(f,100)}\right)^{\frac{1}{(n_h-n_l)}},
\end{equation*}
\item $a_\text{bm}$ is the frequency extrapolated basic median transmission attenuation with respect to free space,  
\item $n_l(\text{h}_\text{BS})$ captures the base station height dependence of the lower distance range power law exponent of the median attenuation relative to free space 
\begin{equation*} 
n_l(\text{h}_\text{BS}) = 0.1(24.9 - 6.55\log{\text{h}_\text{BS}}),
\end{equation*}
\item $n_h$ is the higher distance range power law exponent of the median attenuation relative to free space, 
\item $n$ denotes the modified pathloss exponent
\end{itemize}
\small
\begin{equation*} 
n = 
\begin{cases} 
0.1(24.9 - 6.55\log{\text{h}_\text{BS}}),& \text{if}\, 1\,\text{km} \leq r \leq \text{R}_\text{bp} \\ 
2(3.27\log{\text{h}_\text{BS}} - 0.67(\log{\text{h}_\text{BS}})^2 - 1.75),& \text{if}\, r \geq \text{R}_\text{bp}
\end{cases},
\end{equation*}
\normalsize
\begin{itemize}
\item a($\text{h}_\text{R}$) represents radar reference height correction, and 
\item FSL denotes the free space loss at distance R 
\begin{equation*}
\text{R} = \sqrt{(r\times10^3)^2 + (\text{h}_\text{BS} - \text{h}_\text{R})^2}.
\end{equation*}
\end{itemize}

Typical  fading models such as exponential or Rayleigh model do not accurately capture the observed fading in the link between comms BS and radar. 
This paper models the fading in the link between a comms BS and the radar as a log-normal random variable to capture the effect of large-scale fading due to the relatively large distance between the two. 
Similar reasoning can be applied to model the fading between the radar and its intended target as well, since the latter is typically located tens of kilometers away from the former. 

\subsection{Interference Model}
\label{sec:interf:interf}
Analyzing the aggregate interference at the radar receiver due to comms BS transmissions is an interesting problem, particularly due to the fact that the fading between the two is modeled as a log-normal random variable. The quantity of interest at the radar receiver is the distribution of signal to interference noise ratio (SINR) since the radar target detection performance is characterized by the attained SINR. The SINR at the radar receiver can be modeled as --- 
\begin{equation}
\label{sinr}
\text{SINR} = \frac{\frac{\text{P}_\text{S}\text{G}_\text{T}\Omega\lambda^2\text{F}_\text{R}}{(4\pi)^3\text{R}_\text{T}^4}}{\text{P}_\text{N} + \sum\limits_{k=1}^\text{K} \text{P}_\text{BS}\rho(r_\text{k})\text{F}_\text{BS}g(\phi_k)}
\end{equation}
where 
\begin{itemize}
\item $\text{R}_\text{T}$ is the distance of the target from the radar (radar range), 
\item K denotes the number of comms BS, 
\item $r_\text{k}$ is the distance of $\text{k}^\text{th}$ BS from the radar receiver, 
\item $g(\phi_\text{k})$ is the gain of the radar beam at the $\text{k}^\text{th}$ BS location, and
\item $\text{F}_\text{R}$ and $\text{F}_\text{BS}$ represent the fading random variables between radar-target and BS-radar respectively. 
\end{itemize}

\begin{figure}[t]
\centering 
\includegraphics[width=0.35\textwidth]{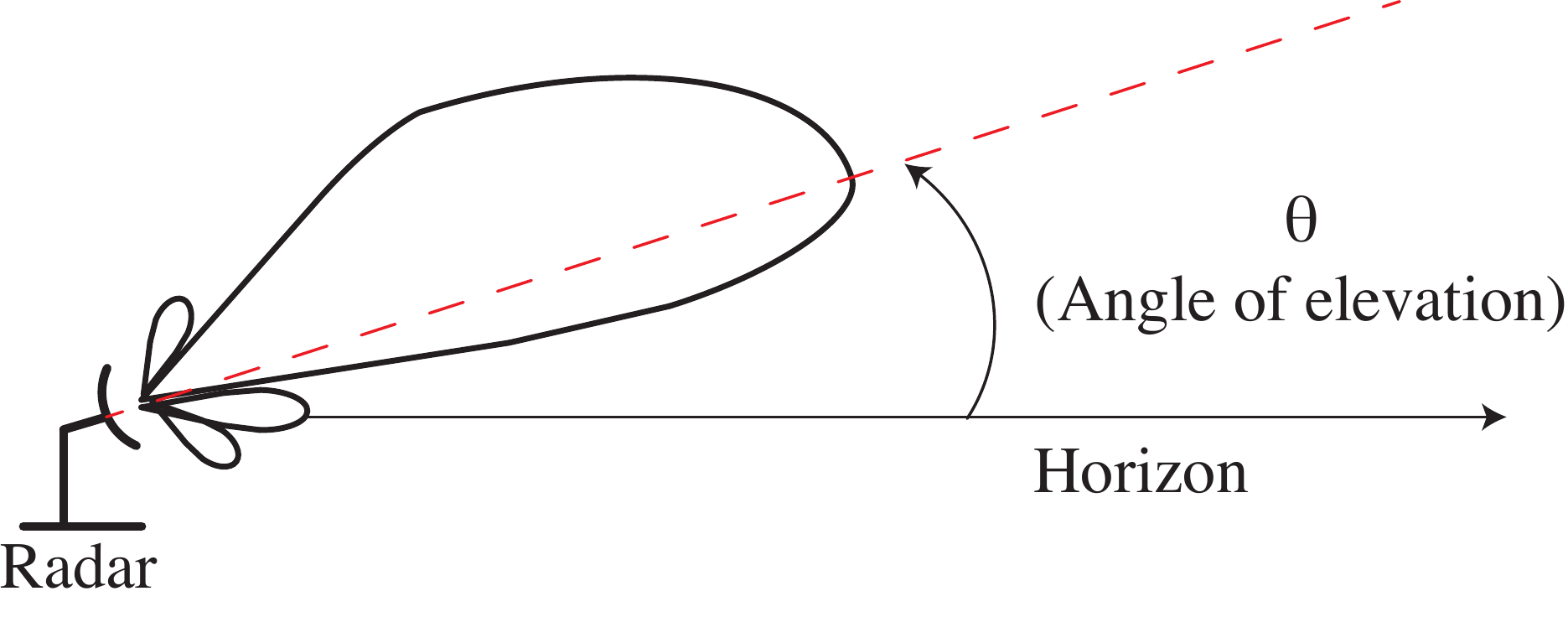}
\caption{Radar with cosecant-square beam pattern operating at an elevation angle of $\theta$ relative to the horizon}
\label{elevation}
\end{figure}

The elevation and azimuth beam patterns of the radar beam are generated based on the model specified in \cite{1851}.
Due to the narrow 3-dB elevation beamwidth of $4.5^\circ$, the angle of elevation of the radar beam plays an important role in modeling the aggregate interference since the beam gain falls steeply within the main lobe. 
A representative diagram describing the elevation angle ($\theta$) of radar beam with respect to the horizon is shown in Fig.~\ref{elevation}. If the radar beam is in the same horizon as the comms BS (that is, if $\theta = 0^\circ$), then interference at the radar receiver from the comms BS would be magnified by the higher beam gain. A discussion on how the elevation angle of radar beam relative to the horizon impacts radar target detection performance is given in section \ref{sec:res:elevation}.


The aggregate interference in the denominator of eq.~\ref{sinr} is a sum of uncorrelated log-normal random variables, which has been analytically characterized by the authors of \cite{lognormal}. Omitting the derivation in the interest of space, the distribution of SINR at the radar receiver can be described as --- 
\begin{equation} 
\label{cdf}
\mathbb{P}[\text{SINR} < \mathcal{T}] = 1-\mathbb{Q}\left(\frac{10\log \mathcal{T}+10\log \Lambda}{\bar{\sigma}}\right)
\end{equation}
where: 
\begin{align*}
\Lambda &= \sum_{k=1}^\text{K} 10^{(t_\text{k} - t_\text{i})/10} + \frac{1}{\gamma} \\ 
t_\text{k} &= \text{P}_\text{BS} \text{(dBm)} - \rho(r_\text{k}) + g(\phi_\text{k}) \\ 
t_\text{i} &= 10\log(\text{P}_\text{R}) - 40\log(\text{R}_\text{T}) \\ 
\text{P}_\text{R} &= \frac{\text{P}_\text{S}\text{G}_\text{T}\Omega\lambda^2}{(4\pi)^3\text{R}_\text{T}^4}, \quad\quad \quad\gamma = \frac{\text{P}_\text{R}}{\text{R}_\text{T}^4\text{P}_\text{N}} \\ 
\bar{\sigma} &= \sigma^2 + \sum_{k=1}^\text{K} \lambda_k^2 \sigma^2 , \quad\lambda_k = \frac{10^{(t_\text{k} - t_\text{i})/10}}{\Lambda}
\end{align*} 
\begin{figure*}[t] 
\centering 
  \subfloat[Radar Range = $20$ km]{\label{range_20}\includegraphics[width=0.32\textwidth]{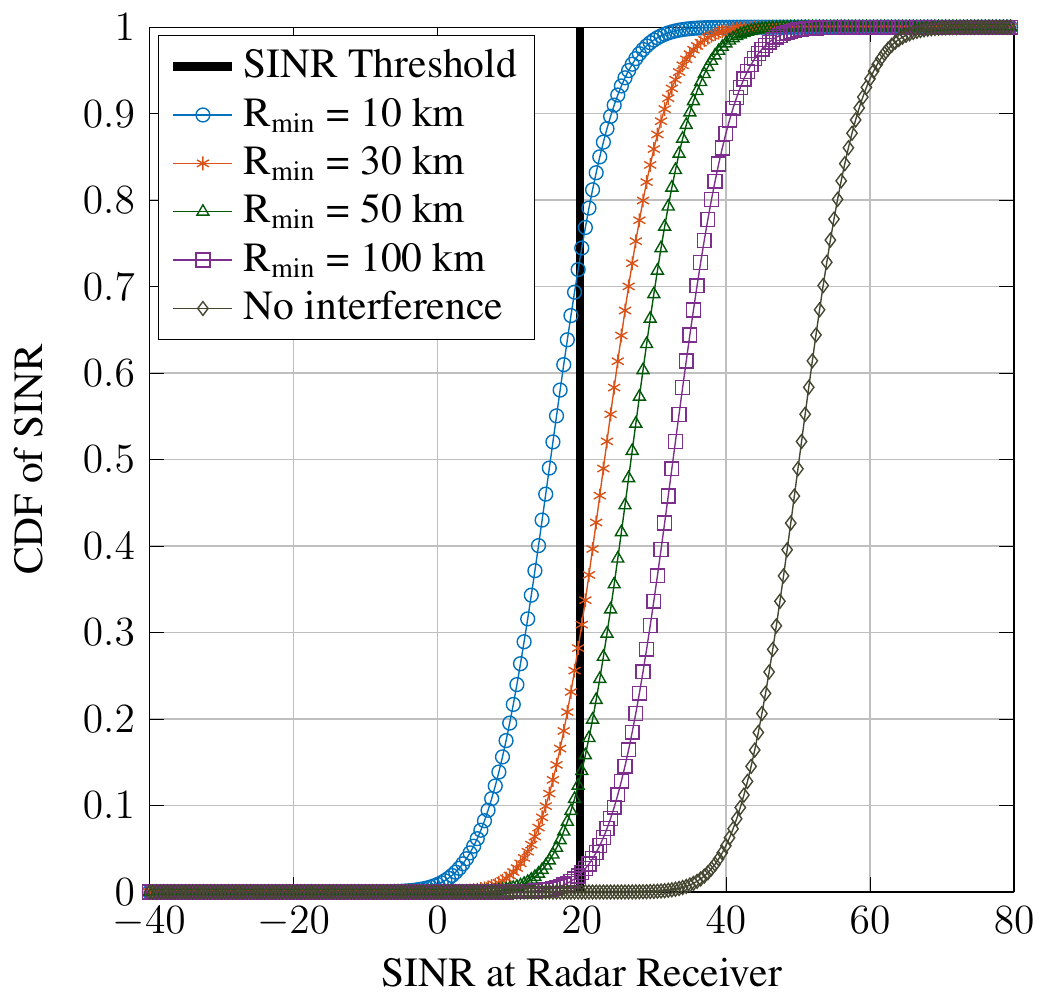}} \hfill
  \subfloat[Radar Range = $60$ km]{\label{range_60}\includegraphics[width=0.32\textwidth]{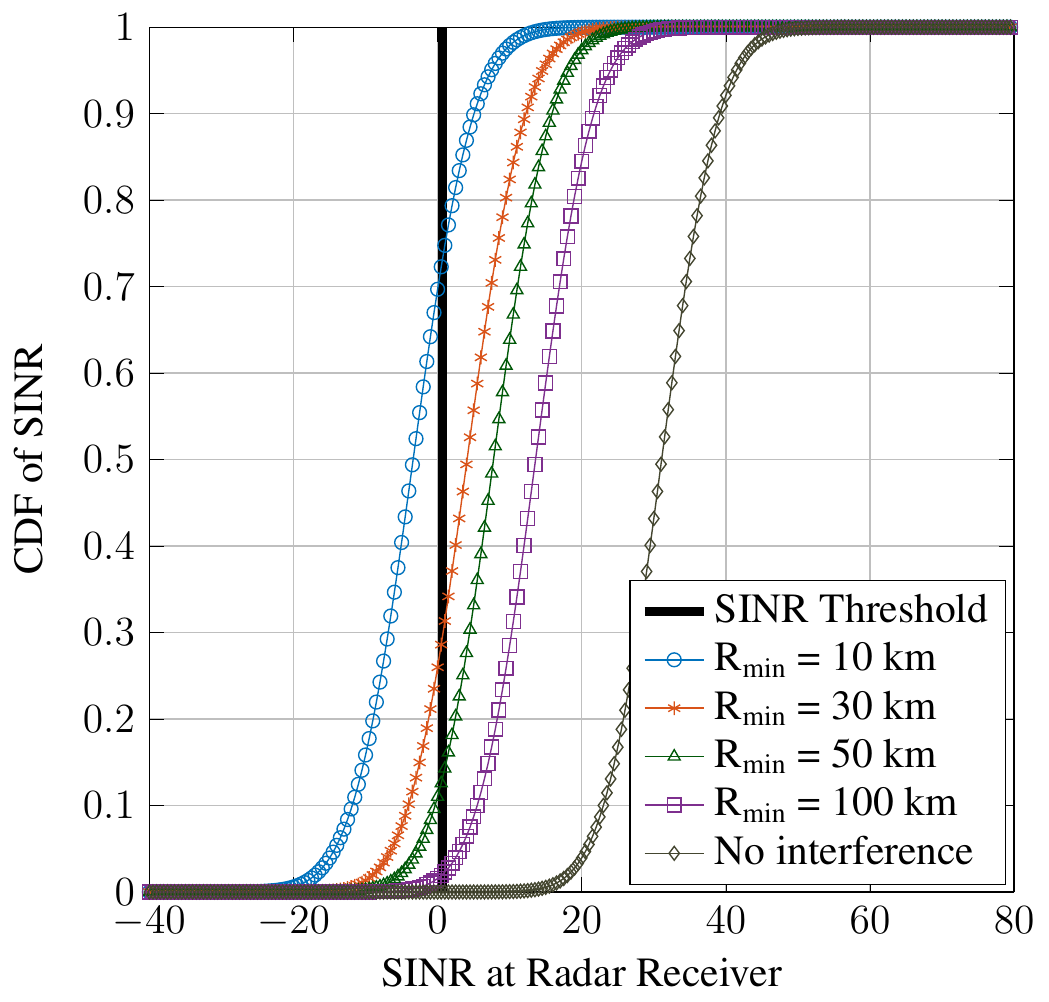}} \hfill
  \subfloat[Radar Range = $80$ km]{\label{range_80}\includegraphics[width=0.32\textwidth]{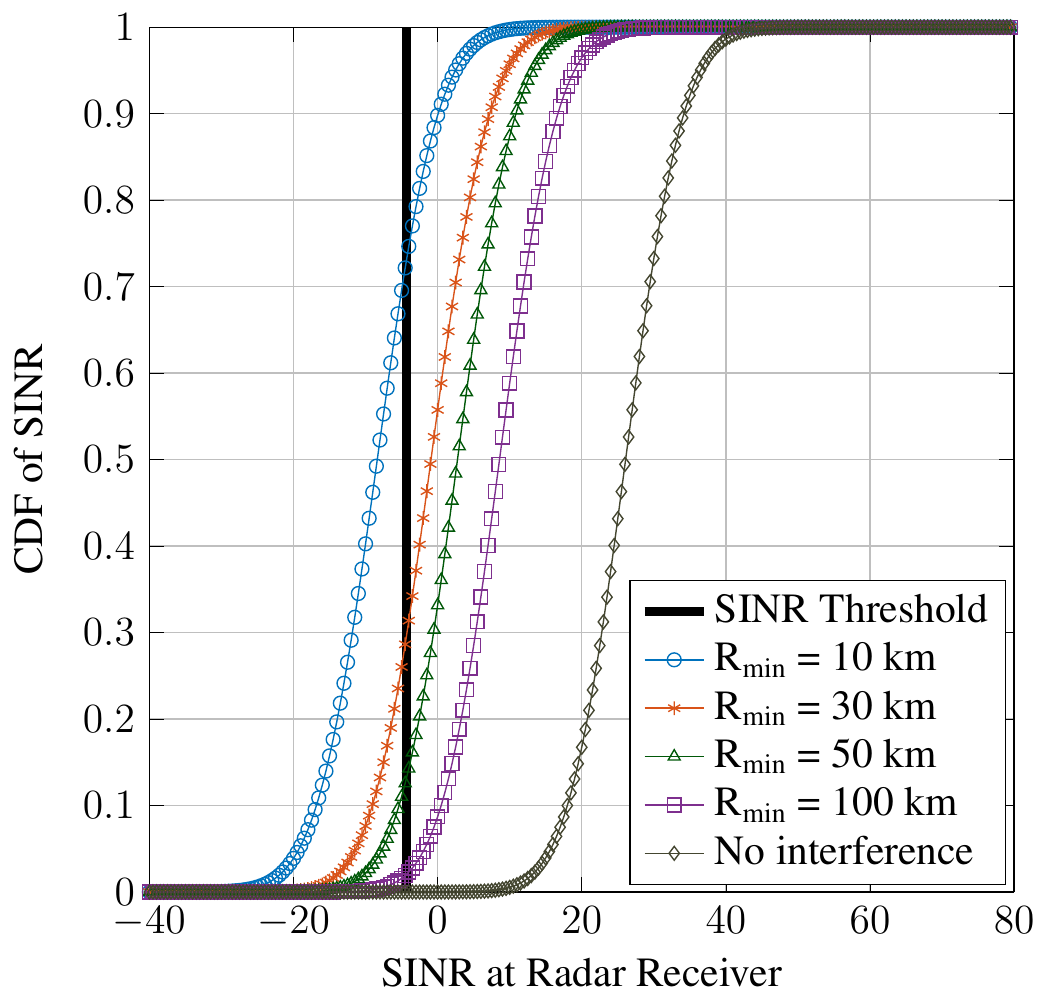}} 
  \centering
  \caption{Distribution of SINR at radar receiver for different radar ranges; protection radius ($\text{R}_\text{min}$) around the radar is the closest distance at which a comms BS may be deployed from the radar}
  \label{plot_sinr}
\end{figure*}
Once the cumulative distribution function (CDF) of SINR from eq.~\ref{cdf} at the radar receiver has been characterized, its detection performance can be quantified in terms of expected probability of target detection ($\mathbb{E}[\text{P}_d]$) as follows --- 
\begin{align}
\label{pd1}
\text{P}_d|_\text{SINR = x} &= \mathbb{Q}(\sqrt{2\times(\text{SINR = x})},\sqrt{\text{P}_{fa}}) \\
\label{pd2}
\mathbb{E}[\text{P}_d] &= \sum_\text{SINR = x} \text{P}_d|_\text{SINR = x}\text{p}_\text{SINR = x}
\end{align}
where 
\begin{itemize}
\item $\mathbb{Q}(\text{a,b})$ is the Marcum's Q function defined as 
\begin{equation*} 
\mathbb{Q}(\text{a,b}) = \int_\text{b}^\infty r \, \text{exp}\left(-\frac{r^2 + \text{a}^2}{2}\right)\,\mathbb{I}_0(\text{a},r)\, \text{d}r, 
\end{equation*}
\item $\mathbb{I}_0(\text{a},r)$ is a modified Bessel function, 
\item $\text{P}_d|_\text{SINR = x}$ is the probability of detection when SINR = x, 
\item $\text{P}_{fa}$ is the desired probability of false alarm, and 
\item $\text{p}_\text{SINR = x}$ is the probability mass function of SINR obtained from its CDF. 
\end{itemize}
\section{NUMERICAL RESULTS AND DISCUSSION}
\label{sec:res:main}
This section first presents the methodology used for obtaining numerical results, and then provides a detailed discussion along with key takeaways based on these results. 

\subsection{Methodology}
\label{sec:res:method}
The operating parameters of radar and comms BS involved in the analysis are listed in Table~\ref{specifications} and \ref{specs_LTE} respectively. 
The interfering comms BS are deployed in an annular region around the radar of interest (as described in Fig.~\ref{scenario}) in such a way that transmissions from one BS reach any other BS at a power less than $-62\, \text{dBm}$ with a probability of $0.9$. This technique is adopted to ensure that comms BS do not deleteriously interfere with each others' transmissions. Monte Carlo simulations are carried out by randomly deploying the comms BS using the aforementioned strategy for different protection radii around radar and the distribution of SINR (eq.~\ref{cdf}) at the radar receiver is averaged out over 10,000 trials. 
In the interest of tractability of analysis, the value of $\text{R}_\text{max}$ is chosen to be $200$~km. 
This is a justifiable assumption as the aggregate interference from comms BS located more than $200$ km away from the radar will be extremely  low at the radar receiver (because of the high pathloss suffered by the transmissions from such comms BS) . After obtaining the averaged-out CDF of SINR, the impact of comms interference on the radar performance is quantified by evaluating the expected probability of target detection of radar (given by eq.~\ref{pd1} and eq.~\ref{pd2}) for different false-alarm probabilities.

In the following discussion of results, the co-channel interfering comms BS are assumed to be in the same horizon as the radar beam, unless explicitly mentioned otherwise. 

\subsection{Distribution of SINR}
\label{sec:res:sinr}
The CDF of SINR at the radar receiver averaged over 10,000 Monte-Carlo runs for different target ranges and protection radii are plotted in Fig.~\ref{plot_sinr}. The fundamental protection criterion that needs to be achieved at the radar receiver, as specified by regulatory authorities, is an interference-to-noise ratio (INR) of at least $\text{INR}_\text{thr}=-10$ dB \cite{ntia2006inr}. 
In other words, the radar can tolerate the comms interference as long as the aggregate interference is not substantial enough to violate the INR protection specification. Therefore, the minimum SINR threshold to be achieved at the radar receiver corresponding to the INR requirement can be computed as ---
\begin{equation}
\label{thr}
\text{SINR}_\text{thr} = \frac{\frac{\text{P}_\text{S}\text{G}_\text{T}\Omega\lambda^2}{(4\pi)^3\text{R}_\text{T}^4}}{(1 + 10^{\text{INR}_\text{thr}/10})\times\text{P}_\text{N}}
\end{equation}
where $\text{P}_\text{N}$ represents the noise power and $\text{R}_\text{T}$ denotes the radar target range. In Fig.~\ref{plot_sinr}, the SINR threshold corresponding to a target range is plotted in bold. If the observed SINR at the radar receiver exceeds $\text{SINR}_\text{thr}$ for a certain target range and protection radius with high probability, then the comms interference to radar in this scenario is deemed to be unobjectionable. In the interest of completeness of analysis, the distribution of SNR at the radar receiver in the absence of any comms interference is also plotted in Fig.~\ref{plot_sinr}. 

It can be seen consistently across different radar ranges in Fig.~\ref{plot_sinr} that the observed SINR at the radar receiver exceeds the required SINR threshold $\sim 70\%$ of the time even when the protection radius is as close as $30$ km. 
This observation holds for any radar range since the aggregate interference effected on the radar by comms BS deployment at a particular protection radius is the same regardless of the radar target range ($\text{R}_\text{T}$),
that is, the aggregate interference plus noise power is independent of $\text{R}_\text{T}$ (denominator of eq.~\ref{thr}). Hence, if the INR protection criterion is satisfied by a comms BS deployment at a certain protection radius, the criterion will be met for any radar range at the same protection radius. 

\begin{figure}[t]
\centering 
\includegraphics[width=0.45\textwidth]{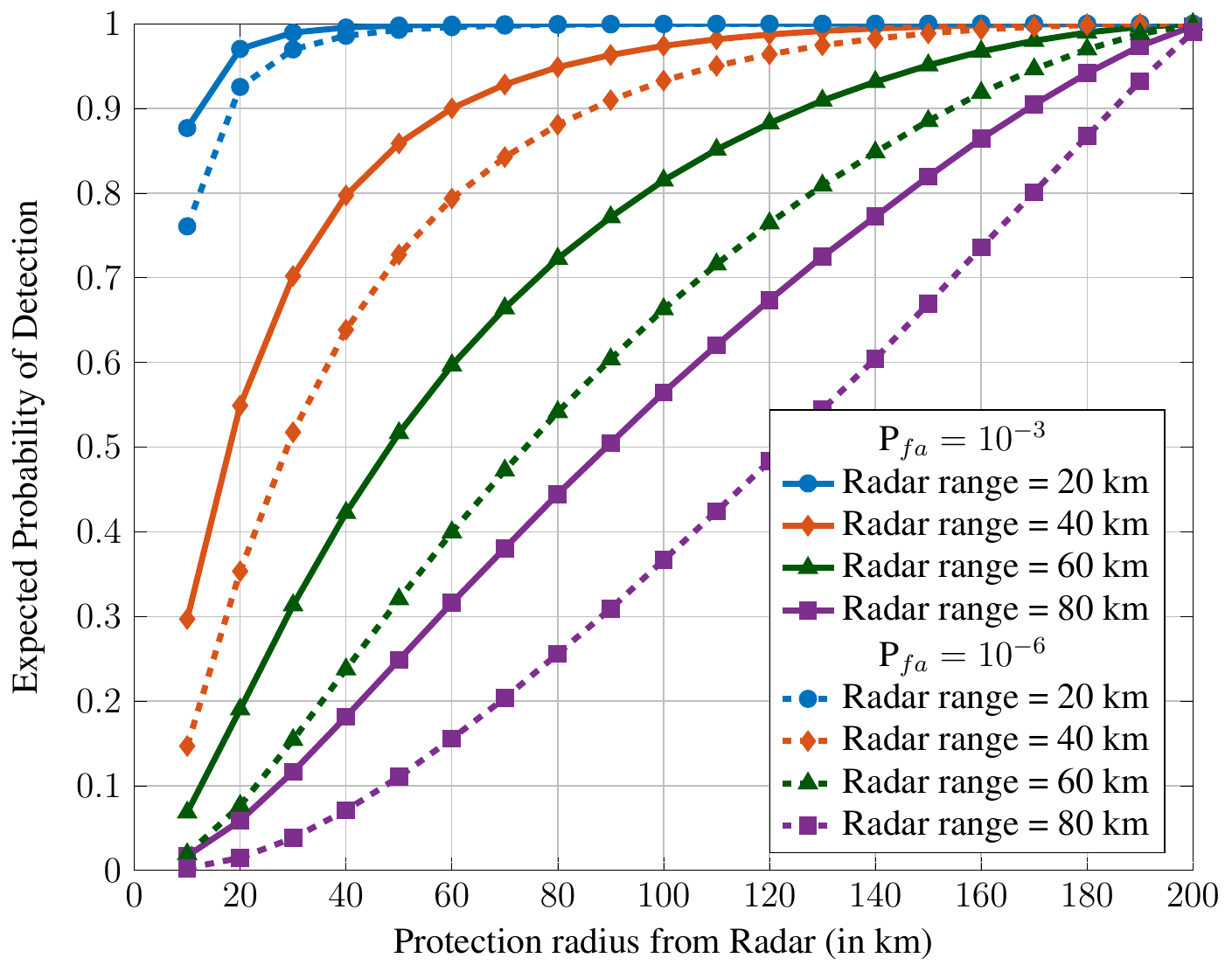}
\caption{Variation of the expected value of probability of target detection achieved by radar for different radar ranges, comms BS protection radii, and false alarm probabilities ($\text{P}_{fa}$) - the radar beam in the same horizon as the interfering comms BS}
\label{plot_pd}
\end{figure}

\begin{figure}[t]
\centering 
\includegraphics[width=0.45\textwidth]{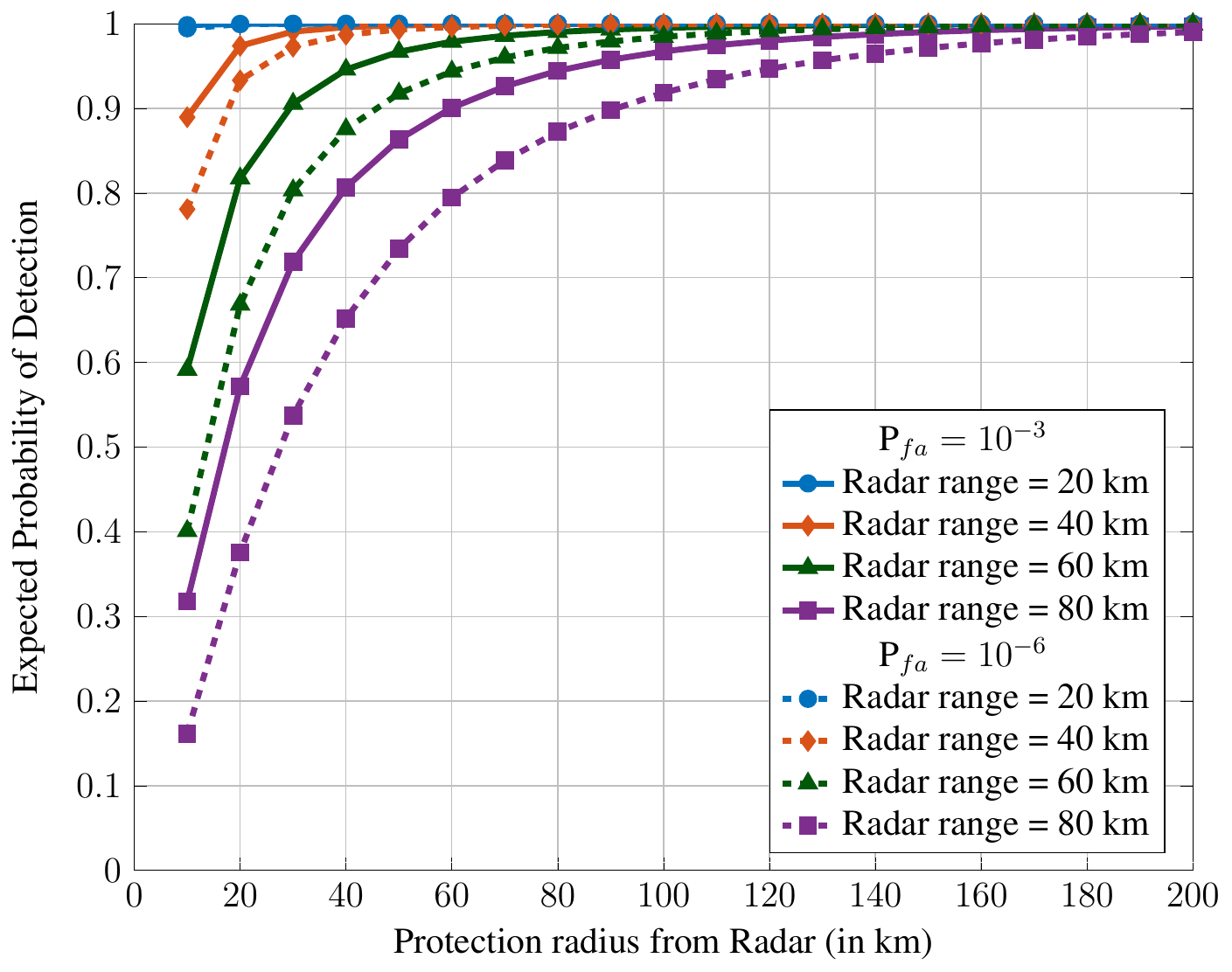}
\caption{Variation of the expected value of probability of target detection achieved by radar for different radar ranges, comms BS protection radii, and false alarm probabilities ($\text{P}_{fa}$) - the radar beam at an elevation angle of $5^{\circ}$ from the horizon}
\label{plot_pd_offset}
\end{figure}

The results in Fig.~\ref{plot_sinr} indicate that if comms BS are deployed with a protection radius of $50$ km  around the radar, the effective interference at the radar receiver will be within tolerable limits more than $90 \%$ of the time (i.e, with probability $0.9$). If the latter requirement (that is the percentage of time aggregate interference is within acceptable threshold) can be relaxed, the protection radius around the radar can be correspondingly shrunk to as low as $30$ km and still result in aggregate interference at the radar that is within acceptable limits more than $70\%$ of the time. Hence, such analysis and observations can provide useful guidelines regarding the choice of protection radius around the radar in order to attain a certain probability that aggregate interference will not exceed the maximum tolerable threshold. 

\subsection{Radar Target Detection Performance}
\label{sec:res:radar}
Although meeting the INR requirement at the radar receiver would `protect' the radar from detrimental interference from the comms system, it is worthwhile to investigate and quantify the impact of comms interference on the radar target detection performance. 
The probability of target detection achieved by a radar is related to the SINR at the radar receiver by eq.~\ref{pd1}. Since this work assumes fading to be random, the distribution of SINR at the radar receiver is characterized and hence the expected value of probability of target detection by radar is computed using the eq.~\ref{pd2}. Note that the probability of target detection is a function of the desired probability of false alarm. 

Fig.~\ref{plot_pd} describes the variation of expected probability of target detection by the radar computed using the CDF of SINR obtained by averaging the Monte Carlo trials for different false alarm probabilities ($\text{P}_{fa}$). When $\text{P}_{fa} = 10^{-6}$, it is observed from Fig.~\ref{plot_pd} that more than $90 \%$ target detection probability is achieved when the radar range is $20$ km and the protection radius is as low as $20$ km. It is also seen that a detection probability of at least $80\%$ can be attained when the radar range is $40$ km by maintaining a protection radius of $60$ km around the radar. Note that the curves approach probability of detection of $1$ with increasing protection radius since the transmissions from interfering comms BS suffer higher pathloss at larger protection radii. To achieve a constant target detection probability across increasing radar ranges, the protection radius has to be correspondingly increased. This is evident from the fact that to maintain a detection probability of $90\%$, a protection radius of $\sim 80$ km around the radar suffices for a radar range of $40$ km but it has to be increased to $\sim 150$ km when the radar range is $60$ km. 

If a false alarm probability greater than $10^{-6}$ is defensible, better probability of target detection is achieved by the radar. As an illustration, consider detection performance of the radar when the radar target range is $40$ km and  protection radius is $30$ km. The radar detects the target just above $50\%$ of the time when $\text{P}_{fa} = 10^{-6}$ but the performance steadily improves as the requirement on $\text{P}_{fa}$ is made less stringent, reaching up to $70\%$ detection probability when a $\text{P}_{fa}$ of up to $10^{-3}$ can be tolerated. Similar improvements are seen in detection performance of targets at different ranges and protection radii. 
\subsection{Effect of Radar Beam Elevation Angle}
\label{sec:res:elevation}
The discussion till now has been under the consideration that the interfering comms BS are on the same horizon as the radar beam, that is, the elevation angle of the radar beam relative to the horizon is $0^\circ$. Due to the narrow 3-dB elevation beamwidth of $4.5^\circ$, the angle of elevation of the radar beam plays an important role in modeling the aggregate comms interference at the radar receiver since the beam gain falls steeply within the main lobe. As an illustration, the variation of expected probability of target detection by the radar when the radar beam is at an elevation angle of $5^\circ$ relative to the horizon is shown in Fig.~\ref{plot_pd_offset}. Comparing the plots in Fig.~\ref{plot_pd} and Fig.~\ref{plot_pd_offset} corresponding to each false alarm probability, it is consistently seen that there is a drastic improvement in target detection performance - with gains as high as $300\%$ in some cases. Thus, if the comms BS are not located in the same horizon as the radar beam, the impact on radar target detection performance is benign compared to when the comms BS are in the same horizon. In other words, if the radar main beam is tilted up relative to the horizon in which the comms BS are located, then the BS can maintain a small protection radius around the radar and still not cause harmful impact on the radar target detection performance. 

\begin{figure}[t]
\centering 
\includegraphics[width=0.45\textwidth]{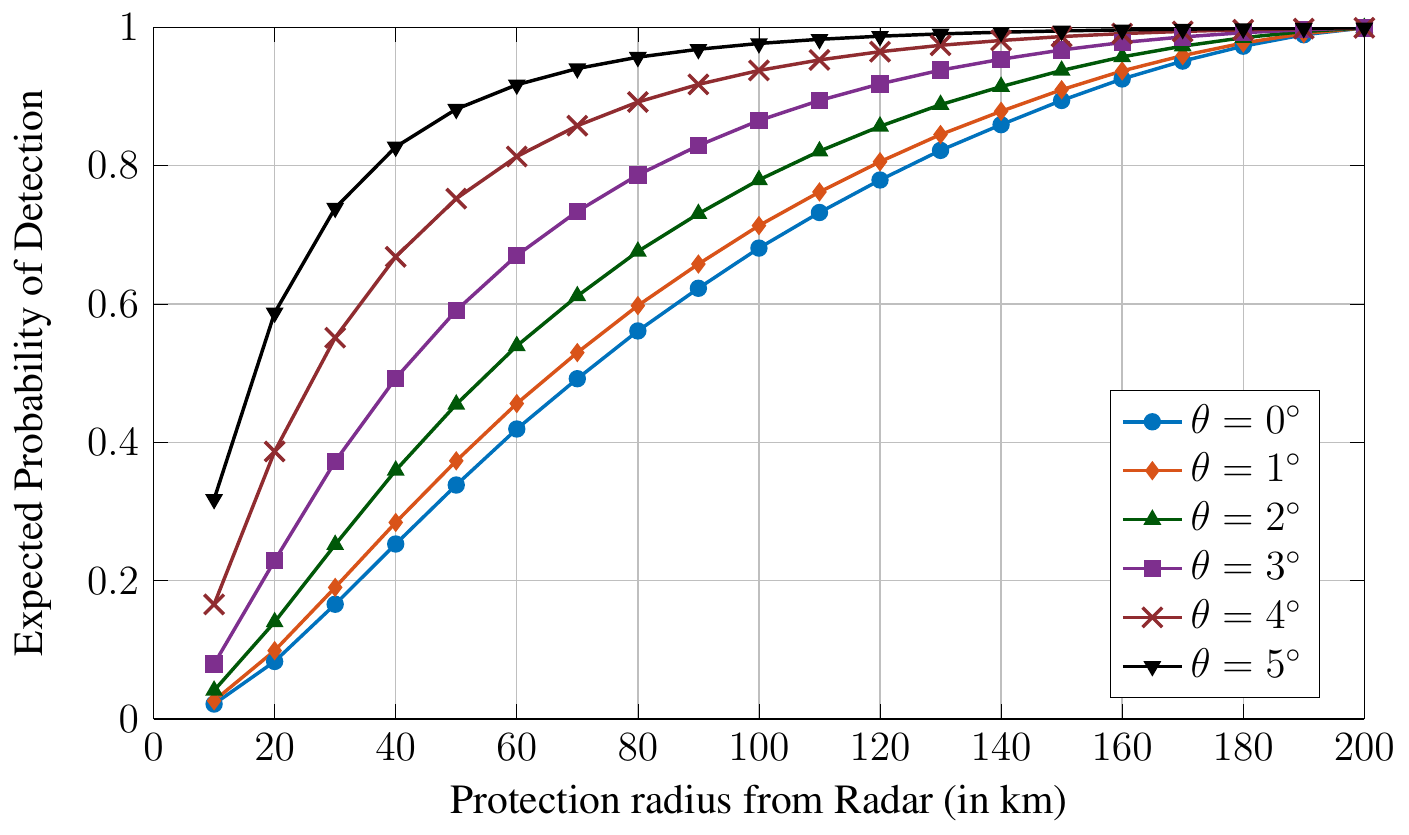}
\caption{Variation of the expected probability of target detection with the elevation angle ($\theta$) of radar beam relative to the horizon; Radar range $\text{R}_\text{T} = 60$ km and desired $\text{P}_{fa} = 10^{-6}$}
\label{fig:plot_pd_diffgains}
\end{figure}

This observation is further corroborated by Fig.~\ref{fig:plot_pd_diffgains} in which the variation of expected probability of target detection by the radar is plotted for different elevation angles of the radar beam with respect to the horizon (as described in Fig.~\ref{elevation}). The radar target range is fixed at $60$ km and the required false alarm probability at $10^{-6}$. It is evident from Fig.~\ref{fig:plot_pd_diffgains} that the radar target detection performance improves appreciably with increasing elevation angle. For instance, to achieve a detection probability of $0.9$ for a radar range of $60$ km, the comms BS have to maintain a protection radius of $\sim 150$ km away from the radar when the beam is in the same horizon as the comms BS but this radius can be shrunk to as low as $50$ km when the elevation angle of the radar beam is $5^\circ$ relative to the horizon. 
\section{CONCLUSION} 
\label{sec:conclude}
This paper investigates the feasibility of coexistence between an air traffic control (ATC) radar, generally stationed in airports, and a wide area wireless communication (comms) system. A comprehensive analysis of interference caused by the comms system on the ATC radar is presented with respect to multiple parameters such as radar range, protection radius, and radar antenna elevation angle relative to the horizon. The analysis suggests that maintaining a protection radius of $50$ km around the radar will guarantee the required INR protection criterion ($=-10$ dB) with comms interference at the radar receiver more than $90\%$ of the time, even when the radar beam is in the same horizon as the interfering comms BS. An evaluation of radar target detection probability indicates that better detection performance is achieved when a higher false alarm probability can be tolerated and/or when the radar main beam is tilted at an elevation angle with respect to the horizon (because of the narrow 3-dB elevation beamwidth of the radar beam). 
These observations suggest that the protection radius around the radar can be trimmed without compromising its detection performance. 
The analysis presented in this paper can hence be used as a general framework to choose a protection radius around any kind of radar to meet a specified detection performance requirement, as long as the radar specifications are known. The choice of ATC radar for the current analysis is essentially because of the easy accessibility of the radar specifications. The results from the current modeling and analysis can also serve as motivation for regulatory bodies to open up bands in the $2700$--$2900$~MHz range for potential unlicensed secondary operation. 
Future directions and extensions of this work may include adaptive power control of the interfering comms BS in order to decrease the protection radius around the radar, and investigating MAC-level scheduling algorithms to facilitate harmonious coexistence between the two systems. 

\section*{ACKNOWLEDGMENT}
This work is supported in part by a grant from the U.S. Office of Naval Research (ONR) under grant number N00014-15-1-2168. The work of S. Kompella is supported directly by the Office of Naval Research.

\bibliographystyle{IEEEtran}
\bibliography{IEEEabrv,milcom} \end{document}